\documentstyle[14pt]{article}

\begin{document}           
\title{
 Report IRB-StP-GR-240696 \newline
\newline \newline \newline
{\bf
The Simplest Exact Solutions in the LTB Model
}
}
\author{\it by \\ \\
{\bf Alexander Gromov}
\\ \\
\small\it St. Petersburg State Technical University \\
\small\it Faculty of Technical Cybernetics, Dept. of Computer Science \\
\small\it 29, Polytechnicheskaya str. St.-Petersburg, 195251, Russia \\
\small and \\
\small\it Istituto per la Ricerca di Base \\
\small\it Castello Principe Pignatelli del Comune di Monteroduni \\
\small\it I-86075 Monteroduni(IS), Molise, Italia \\
\small\it e-mail: gromov@natus.stud.pu.ru
}
\date{}
\maketitle
\begin{abstract}

The rules of calculating three undetermined functions which defined a
solution in the LTB model are used to study the class of exact
nonhomogene\-ous models with $f^2(\mu) = 1$, $\Lambda = 0$.
The parameter $\nu(\mu)$ defined the
difference between LTB and FRW models is found out and
the limit transformation to the FRW model is shown.
The initial conditions are present throught density and Habble function
at the moment of time $\tau = 0$.
Two criteria of homogeneous of matter distribution are studied.
The asimptotic of the present solution for  $\tau \rightarrow
+\infty$ is studied.
\\
PACS number(s):98.80
\end{abstract}
\newpage
\section{The Introduction} \label{introd}

This article is dedicated to studying of
the simplest solution in the LTB model and
based on the results are obtained in \cite{Tolman:34},\cite{Bonnor:72} and
\cite{A.G.1} and presented shortly in this section.
The LTB model is studied here as the Cauchy problem for the PDE
\cite{Tolman:34}:
\begin{equation}
e^{3 \,\omega(\mu,\tau) / 2}
\left(
\frac{\dot \omega^2(\mu,\tau)}{2} - \frac{2}{3} \Lambda
\right)
+2 e^{\omega(\mu,\tau) /2} \left[1 - f^2(\mu)\right] = F(\mu)
\label{T:11}
\end{equation}
whit initial conditions
\begin{equation}
\left.
\begin{array}{c}
\left.\omega(\mu,\tau)\right|_{\tau=0} = \omega_0(\mu), \quad
\left.\dot\omega(\mu,\tau)\right|_{\tau=0} = \dot\omega_0(\mu), \quad \\ \\
\left.\ddot\omega(\mu,\tau)\right|_{\tau=0} = \ddot\omega_0(\mu)
\end{array}
\right\}
\label{init}
\end{equation}
and constants
\begin{eqnarray}
\left.
\begin{array}{c}
\left.\omega(\mu,0)\right|_{\mu=0} = \omega_0(0), \quad
\left.\dot\omega(\mu,0)\right|_{\mu=0} = \dot\omega_0(0), \quad \\ \\
\left.\ddot\omega(\mu,0)\right|_{\mu=0} = \ddot\omega_0(0), \quad
\Lambda,
\end{array}
\right\}
\label{init-c}
\end{eqnarray}
where $\mu$ and $\tau$ are dimensionless co-moving coordinates,
corresponding to Lagrangian coordinate and time,
$\omega(\mu,\tau)$ is the metrical function,
$\dot{} = \frac{\partial}{\partial \tau}$, $\Lambda$ is the cosmological
constant,
\begin{equation}
f^2(\mu) - 1 =
e^{\omega_0(\mu)}
\left(
\ddot \omega_0(\mu) + \frac{3}{4} \dot \omega^2_0(\mu) - \Lambda
\right),
\label{DEF:f^2-1}
\end{equation}
\begin{equation}
F{\rm(\mu)} =
e^{3 \,\omega_0(\mu) / 2}
\left(
\frac{\dot \omega^2_0(\mu)}{2} - \frac{2}{3} \Lambda
\right)
+2
e^{\omega_0(\mu) /2}
\left[1 - f^2(\mu)\right].
\label{DEF:itF}
\end{equation}
The law of the density has the form
\begin{eqnarray}
{\rm
    8 \pi \delta(\mu,\tau) =
    \frac{e^{
             \frac{3}{2}
             [\omega_0(\mu) - \omega(\mu,\tau)]
            }
         }
         {
          \omega^{\prime}(\mu,\tau)
         }
} \times
\nonumber\\
{\rm
    \left\{
           3\left[\omega_0(\mu)\right]^{\prime}
            \left[
                  -\ddot\omega_0(\mu) - \frac{1}{2}\dot\omega_0^2(\mu) +
                  \frac{\Lambda}{6}
            \right]
     - 2\left[\ddot\omega_0(\mu)\right]^{\prime} -
     2\dot\omega_0(\mu)\left[\dot\omega_0(\mu)\right]^{\prime}
    \right\}
}
\label{DEF:rho_new}
\end{eqnarray}
The Habbl function is defined by the expression
\begin{equation}
h(\mu,\tau) = \frac{\dot R^{\prime}(\mu,\tau)}{R^{\prime}(\mu,\tau)}
= \frac{\partial \ln{R^{\prime}(\mu,\tau)}}{\partial \tau}.
\label{Habble_def}
\end{equation}
This article studies the simplest solution in the LTB model defining by
the initial conditions
\begin{equation}
f^2(\mu) = 1, \qquad \Lambda = 0.
\label{a}
\end{equation}
For the first time this solution has obtained by Bonnor \cite{Bonnor:72}.
The Bonnor's result is wrote for the Bonnor's function
\begin{eqnarray}
R(\mu,\tau) = e^{\omega(\mu,\tau)/2},
\nonumber
\end{eqnarray}
which has the sense of Euler coordinate \cite{L&L}, \cite{LE&AD}:
\begin{eqnarray}
R(\mu,\tau) = \frac{9^{1/3}}{2}\left[
\tau \sqrt{\Psi(\mu)} + \beta(\mu)
\right]^{2/3},
\label{B:1}
\end{eqnarray}
where $\Psi(\mu)$ and $\beta(\mu)$ are arbitrary functions.

This article is devoted to study the same problem through the metrical
function and initial conditions for it.

\section{The $f^2(\mu) = 1$, $\Lambda = 0$ Solution} \label{solution}

The initial conditions (\ref{a}) generate the simplest analytical
solution.
The condition $f^2(\mu) = 1$, as it goes from (\ref{DEF:f^2-1}),
correlates the initial conditions as follows:
\begin{equation}
\ddot \omega_0(\mu) + \frac{3}{4} \dot \omega^2_0(\mu) = 0.
\label{init_fa}
\end{equation}
Due to this reason,
the general number of the initial conditions (\ref{init}) is decreased by
one unit.

We start with the solution of the equation
(\ref{T:11}).
It is follows from
(\ref{T:11}) that:
\begin{equation}
\left(
\frac{\partial e^{\omega(\mu,\tau)/2}}{\partial \tau}
\right)^2 =
f^2(\mu) - 1 + \frac{1}{2} F(\mu) e^{-\omega(\mu,\tau)/2} +
\frac{\Lambda}{3} e^{\omega}.
\label{dif_1}
\end{equation}
In the general case the function $F(\mu)$ is defined by the equation
(\ref{DEF:itF}).
Whit help of (\ref{a}) and (\ref{init_fa}) it takes the form
\begin{eqnarray}
F(\mu) = \frac{\dot\omega_0^2(\mu)}{2}e^{3\omega_0(\mu)/2}
\nonumber
\end{eqnarray}
and the equation (\ref{dif_1}) with conditions (\ref{a}) becomes as
follows:
\begin{equation}
e^{3\omega_0(\mu)/4}\frac{
\partial e^{3\omega(\mu,\tau)/4}}{\partial \tau}
= \pm\frac{3}{4}\dot\omega_0(\mu).
\label{solution_1}
\end{equation}
We obtain by the integration of (\ref{solution_1})
\begin{eqnarray}
e^{\frac{3}{4}[\omega(\mu,\tau) - \omega_0(\mu)]} =
\pm\frac{3}{4}\dot\omega_0(\mu)\tau + {\bf F}(\mu).
\nonumber
\end{eqnarray}
The function ${\bf F}(\mu)$ is found out from the initial conditions then
time $\tau = 0$:
\begin{eqnarray}
{\bf F}(\mu) = 1.
\nonumber
\end{eqnarray}
The solution now is:
\begin{equation}
e^{\frac{3}{4}[\omega(\mu,\tau) - \omega_0(\mu)]} = 1
\pm\frac{3}{4}\dot\omega_0(\mu)\tau.
\label{solution_3}
\end{equation}
In the Bonnor's notation the equation (\ref{solution_3}) is readied:
\begin{eqnarray}
\frac{R(\mu,\tau)}{R_0(\mu)} = \left(
1 \pm\frac{3}{4}\dot\omega_0(\mu)\tau
\right)^{2/3},
\label{DEF:F_f_1}
\end{eqnarray}
where the upper sign correspond to the expansion from the initial condition
into the infinity and lower sign correspond to the collapse from one.

The collapse solutions contains the characteristic time of collapse
depending on the initial
mass coordinate: if the particle has the coordinate $\mu$ at the
moment $\tau = 0$, the time of collapse is equal to
\begin{equation}
\bar\tau(\mu) = \frac{4}{3\dot\omega_0(\mu)}.
\label{time}
\end{equation}
>From this moment and up to the end of this section the dependence of
$\omega$ from $\mu$ and $\tau$ will be omitted.
We obtain the density corresponding to this solution by the substitution
(\ref{solution_3}) into (\ref{DEF:rho_new}). Let's denote
\begin{equation}
\nu = \frac{[\dot\omega_0]^{\prime} }{ [\omega_0]^{\prime} }.
\label{nu-def}
\end{equation}
So, the density is:
\begin{equation}
8 \pi \delta(\mu,\tau) =
\displaystyle
\frac{\dot\omega_0}{\left(
1 \pm\frac{3}{4}\dot\omega_0\tau
\right)^2} \ \cdot
\displaystyle
\frac{
\nu + \frac{3}{4} \dot\omega_0
}
{
1 \pm
\displaystyle
\frac{\nu\tau}{
1 \pm\frac{3}{4}\dot\omega_0\tau
}
},
\label{ro-sol}
\end{equation}

Using the definition (\ref{Habble_def}) we find out the Habble's function
for the solution (\ref{solution_3}):
\begin{eqnarray}
h(\mu,\tau) =
\pm\frac{\frac{3}{4}\dot\omega_0 + \nu}
            {1 \pm \left(\frac{3}{4}\dot\omega_0 +\nu\right)\tau}
\mp
\frac{1}{4} \, \cdot
\frac{\dot\omega_0}
     {1 \pm \frac{3}{4}\dot\omega_0\tau}
\label{Habbl-sol}
\end{eqnarray}
Knowing the law of the density and Habble's function we obtain now
the formulum for the cosmological parameter $\Omega$:
\begin{eqnarray}
\Omega = \frac{\rho}{\rho_c}, \quad \mbox {where} \quad
\rho_c = \frac{3H_0^2}{8\pi G},
\nonumber
\end{eqnarray}
where $H_0$ and $\rho$ mean the dimension Habble function and the density
at the moment of the observation.
Let's assume that this moment is $\tau = 0$.
The density, the critical density and Habble function at this moment are:
\begin{equation}
8 \pi \delta(\mu,0) =
\dot\omega_0 \left(\frac{3}{4}\dot\omega_0 + \nu\right),
\label{ro-sol-a}
\end{equation}
\begin{eqnarray}
\delta_{c}(\mu,0) = \frac{3 \tilde{H}^2}{8 \pi G \rho(0,0)} h^2(\mu,0),
\nonumber
\end{eqnarray}
\begin{equation}
h(\mu,0) = \pm\left(\frac{1}{2}\dot\omega_0 +\nu\right),
\label{Habbl-sol-12}
\end{equation}
\begin{eqnarray}
\Omega(\mu,0) =
\frac{G \rho_0}{3 \tilde{H}^2}
\dot\omega_0\frac{\frac{3}{4}\dot\omega_0 + \nu}{\left(
\frac{\dot\omega_0}{2} + \nu
\right)^2}.
\nonumber
\end{eqnarray}
These functions dependent on time as follow:
\begin{eqnarray}
\delta_{c}(\mu,\tau) = \frac{3 \tilde{H}^2}{8 \pi G \rho_0}
\left[
\pm\frac{\frac{3}{4}\dot\omega_0 + \nu}
            {1 \pm \left(\frac{3}{4}\dot\omega_0 +\nu\right)\tau}
\mp
\frac{1}{4} \, \cdot
\frac{\dot\omega_0}
     {1 \pm \frac{3}{4}\dot\omega_0\tau}
\right]^2,
\nonumber
\end{eqnarray}
\newpage
$$
\Omega(\mu,\tau) =
\frac{16 G \rho_0}{3 \tilde{H}^2} \,
\dot\omega_0\left(\frac{3}{4}\dot\omega_0 +
\nu\right) \, \times
$$
\begin{eqnarray}
\frac{
      \left(1 \pm \frac{3}{4}\dot\omega_0\tau\right)
      \left[1 \pm \left(
      \frac{3}{4}\dot\omega_0 + \nu
      \right) \tau \right]
     }
     {
     \left\{
            \pm 4 \left(
                        1 \pm \frac{3}{4}\dot\omega_0\tau
                   \right)
            \left(
                  \frac{3}{4}\dot\omega_0 + \nu
            \right)
     \mp \dot\omega_0
     \left[
           1 \pm
           \left(
                 \frac{3}{4}\dot\omega_0+\nu
           \right) \tau
     \right]
     \right\}^2
     }\qquad .
\nonumber
\end{eqnarray}
%


\section{The FRW Model} \label{nabl}

Every nonhomogeneous solution of the LTB model should definitely
include the FRW model as a limited case
when the density and Habble function are
not dependent on space coordinate for
all moments of time.
Let's study in which case the present solution is
reduced to the FRW model?
Only one condition satisfys this request:
\begin{equation}
\omega_0(\mu) = const \qquad \mbox{so,} \quad \nu(\mu) = 0.
\label{vu = 0}
\end{equation}
It goes from (\ref{ro-sol}), (\ref{Habbl-sol})  and (\ref{vu = 0}) that in
this case
\begin{equation}
8 \pi \delta(\tau) = \frac{3}{4} \left(
\frac{\dot\omega_0}{1 \pm \frac{3}{4}\dot\omega_0\tau}
\right)^2,
\label{rho_1_2}
\end{equation}
\begin{eqnarray}
h(\tau) = \pm\frac{1}{2} \, \cdot
\frac{\dot\omega_0}{1 \pm\frac{3}{4}\dot\omega_0\tau}.
\nonumber
\end{eqnarray}
>From these equations it goes:
\begin{eqnarray}
8 \pi \delta(\tau) = 3 h^2(\tau).
\label{Habbl_1_3}
\end{eqnarray}
We obtain the meaning of the $\omega_0$ by substitution $\tau = 0$ in
(\ref{rho_1_2}):
\begin{equation}
\dot\omega_0 = \pm 2 \sqrt{\frac{8 \pi}{3}}.
\label{chislo}
\end{equation}
For the collapse solution, in accardence with (\ref{time}) all matter fall
down to the centre simultaneouly at time
\begin{equation}
\bar\tau(\mu) = \frac{1}{\sqrt{6\pi}}.
\label{time-1}
\end{equation}
$\nu(\mu) = 0$ together with (\ref{chislo}) give
\begin{eqnarray}
\Omega = \frac{G \rho_0}{\tilde{H}^2}.
\nonumber
\end{eqnarray}

\section{The Initial Conditions and Criteria of Homogeneous}

We are able now to construct here the initial conditions for the $f^2 = 1$,
$\Lambda = 0$ solution. To do this, we assume the functions $\delta(\mu,0)$
and $h(\mu,0)$ are given and express the initial conditions $\omega_0(\mu)$
and $\dot\omega_0(\mu)$ through $\delta(\mu,0)$ and $h(\mu,0)$.
>From (\ref{ro-sol-a}) and (\ref{Habbl-sol-12}) we find out:
\begin{eqnarray}
\dot\omega_0(\mu) = 2 \epsilon h(\mu,0) \pm
\sqrt{4 h^2(\mu,0) + 32 \pi \delta(\mu,0)}
\label{nu-omega}
\end{eqnarray}
and
\begin{eqnarray}
\nu(\mu) =
\frac{8 \pi \delta(\mu,0)}{
2 \epsilon h(\mu,0) \pm \sqrt{4 h^2(\mu,0) + 32 \pi \delta(\mu,0)}} -
\nonumber\\ \nonumber\\
\frac{3}{4}
\{
2 \epsilon h(\mu,0) \pm \sqrt{4 h^2(\mu,0) + 32 \pi \delta(\mu,0)}
\}
\label{nu-nu}
\end{eqnarray}
where $\epsilon = -1$ for the expansion solution and $\epsilon = +1$
for the collapse one.
It was shown in the section (\ref{nabl}) that the general solution is
reduced to the FRW by the equality $\nu(\mu) = 0$.
Solving the equation $\displaystyle{\nu(\mu) = 0}$ with $\nu(\mu)$ from
(\ref{nu-nu}) we obtain for $\displaystyle\frac{8\pi \delta}{h^2}$ the
quadratic equation which has two solutions:
\begin{eqnarray}
\frac{8\pi \delta}{h^2} = 0 \qquad \mbox{É} \qquad
\frac{8\pi \delta}{h^2} = 3.
\nonumber
\end{eqnarray}
First of them is the solution for the empty space and the second one
is the solution for the FRW model.

We will study now two criteria of homogeneously of the matter distribution.
First of them is the difference
\begin{eqnarray}
8 \pi \delta(\mu,0) - 3 h^2(\mu,0) = -\nu(\mu) \left[
2\dot\omega_0(\mu) + 3 \nu(\mu)
\right]
\label{difer}
\end{eqnarray}
The right part of this equation is equal to zero in two cases:
\begin{equation}
\nu(\mu) =
 \frac{
       [\dot\omega_0(\mu)]^{\prime}
      }
      {
       [\omega_0(\mu)]^{\prime}
      } = 0
\label{case-1}
\end{equation}
and
\begin{eqnarray}
\nu(\mu) = \frac{[\dot\omega_0(\mu)]^{\prime}}{
[\omega_0(\mu)]^{\prime}} = - \frac{2}{3}\dot\omega_0(\mu).
\label{case-2}
\end{eqnarray}
Solving the equation (\ref{case-2}) as differential equation for the
function $\dot\omega_0(\mu)$ (initial conditions)
we find out the correlation
$\dot\omega_0(\mu)$ and $\omega_0(\mu)$:
\begin{eqnarray}
\dot\omega_0(\mu) = const \,\cdot e^{2\omega_0(\mu)/3}
\nonumber
\end{eqnarray}
But the solution (\ref{case-2}) do not assume the density $\delta(\mu,0) =
const$
excluding the case (\ref{case-1}).This example shows the existence of such
initial conditions that the equality (\ref{difer}) is satisfied for density
not equal to constant at the moment of time $\tau = 0$.

The second criterium is the ratio
$\displaystyle\frac{\delta_{LTB}}{\delta_{FRW}}$ where $\delta_{LTB}$ is
from (\ref{ro-sol}) and $\delta_{FRW}$ is from (\ref{rho_1_2}):
\begin{eqnarray}
\frac{\delta_{LTB}}{\delta_{FRW}} =
\frac{4}{3\dot\omega_0}\left(
1 \pm \frac{3}{4}\dot\omega_0\tau
\right)
\frac{
\frac{3}{4}\dot\omega_0 + \nu
}{
1 \pm \left(
\frac{3}{4}\dot\omega_0 + \nu
\right)\tau
}
\label{kr-1}
\end{eqnarray}
In accordance with the FRW solution the equality $\nu(\mu) = 0$ implies the
equality $\displaystyle\frac{\delta_{LTB}}{\delta_{FRW}} = 1$.

\section{The Property of the Solution for \newline $\tau \rightarrow
+\infty$}

Let's study now the expansion nonhomogeneous model with
presented in the section (\ref{solution}) for time
\begin{eqnarray}
\tau \gg \bar\tau(\mu) = \frac{4}{3\dot\omega_0(\mu)}.
\label{area}
\end{eqnarray}
The sense of time $\bar\tau$ in shown by (\ref{time}) for collapse
solution.
In the expansion model it is characteristic value connected with initial
conditions: $\mu$ is the initial position of the particle at the moment of
time $\tau = 0$.
In the limit (\ref{area}) the density and Habble's function are not
dependent on $\mu$-coordinate but only on time:
\begin{eqnarray}
8\pi \delta(\tau) \rightarrow \frac{4}{3\tau^2},
\nonumber
\end{eqnarray}
\begin{eqnarray}
h(\tau) \rightarrow \frac{2}{3\tau},
\nonumber
\end{eqnarray}
so, the follows equation takes a place in the limit (\ref{area}):
\begin{eqnarray}
8\pi\cdot \lim_{\tau \gg \bar\tau(\mu)} \delta(\tau) =
3 \cdot \lim_{\tau \gg \bar\tau(\mu)} h^2(\tau)
\nonumber
\end{eqnarray}
This property allows to name the solution
(\ref{solution_3})  - (\ref{DEF:F_f_1})
of the nonhomogeneous model in
the area (\ref{area}) by FRW-limit of LTB model.
The function $\Omega(\mu,\tau)$ has the form in this limit:
\begin{eqnarray}
\Omega(\mu,\tau) \rightarrow 3 \Omega_0, \qquad \mbox{where} \qquad
\Omega_0 = \frac{8 \pi G \rho_0}{3 \tilde H^2}.
\nonumber
\end{eqnarray}
The function $\Omega(\mu,\tau) = const$ in the limit (\ref{area}) and
it is in accordance with condition $f^2(\mu) = 1$ for the using solution.

\section{Resultes and Conclusions}

This article is dedicated to the development of the Bonnor's solution
(\ref{B:1}) on the grounde of the results has been obtained in
\cite{A.G.1}.
The simplest nonhomogeneous LTB model (\ref{solution_3})  -
(\ref{DEF:F_f_1}) generated by the initial conditions (\ref{a}) and
(\ref{init_fa}) is studied and the follow results are obtained:
\begin{itemize}
\item The time of collapse (\ref{time}) definded by the initial condition
$\dot\omega_0(\mu)$;
\item The expression of the initial conditions through density and Habble
function given at time $\tau = 0$ are persented in (\ref{nu-omega}) and
(\ref{nu-nu});
\item The example of the initial condition with density
not dependent on cordinate $\mu$,
but sutisfyed the FRW equation (\ref{Habbl_1_3}) is shown.
The criteria of homogeneouse of the matter distribution in the present
solution is shown in (\ref{kr-1});
\item It is shown that the LTB solution (\ref{solution_3})  -
(\ref{DEF:F_f_1}) strives for FRW for time from (\ref{area});
\item The parameter $\nu(\mu)$ from (\ref{nu-nu}) definded by the initial
conditions $\omega_0(\mu)$ and $\dot\omega_0(\mu)$ shows the difference
between LTB and FRW models;
\item The particulary case of initial conditions $\dot\omega_0(\mu) =
const$, $\nu(\mu) = 0$ generat the FRW model;
\item The time of collapse in the FRW model is not dependent on $\mu$, so
all matter fall down to the centre simultaneouly.
\end{itemize}

\section{Acknowledgements}

I'm grateful to Prof.Arthur D.Chernin and Dr.Yurij Barishev for
encouragement and discussion.
This paper was financially supported by "COSMION" Ltd., Moscow.

\end{document}